# NOTE ON THE DIRAC FIELD IN REAL DOMAIN

S.BOTRIĆ

Faculty of Electrical Engineering,Mechanical Engineering and
Naval Architecture,University of Split,Split,Croatia

Abstract

Sufficient conditions for linear electrodynamics with real Dirac field are derived.

## 1. INTRODUCTION

Theory of real Dirac field interacting with the electromagnetic field has been put forward in the article [1].Depending on the solutions of Lagrange's and canonical equations of the real Dirac field one may establish either linear electrodynamics with the real Dirac field or nonlinear electrodynamics with the real Dirac field.Only those solutions of canonical equations that represent the origin of linear electrodynamics are concerned in this paper.

## 2. REAL DIRAC FIELD WITH THE ELECTROMAGNETIC INTERACTION

The Lagrange's variable of the real Dirac field $\Phi$ is an eight-component spinor matrix $\Phi$,

$$\Phi = \begin{bmatrix} \Phi_1 \\ \Phi_2 \\ \Phi_3 \\ \Phi_4 \\ \Phi_5 \\ \Phi_6 \\ \Phi_7 \\ \Phi_8 \end{bmatrix}$$

,where $\Phi_i$ (i=1,2,...,8) are real field functions and the four-vectors $A^\alpha$ and $C^\beta$ are the Lagrange's variables of the electromagnetic field.

The Lagrangian density that defines the real Dirac field with the electromagnetic interaction is given by [1]

$$L_{DI} = K\left[\frac{\overline{\Psi}(1-a)\Psi}{1-a^2} - \kappa^2 \Phi(1+a)\Phi\right]$$

(1)

where $a = \dfrac{e}{K}A_\mu \eta^\mu$, $K = mc^2$, $\kappa = \dfrac{mc}{\hbar}$, e is a given point charge, (2)

$\Psi = D\Phi$, $D = \partial_\alpha \eta^\alpha$, (3)

and $\eta^\mu$ are square matrices defined by

$$\eta^0 = \begin{bmatrix} 1 & 0 \\ 0 & 1 \end{bmatrix}_{8\times 8} \quad , \quad \eta^i = \begin{bmatrix} 0 & a^i \\ -a^{i+} & 0 \end{bmatrix}_{8\times 8} \quad ,$$

$$a^1 = \begin{bmatrix} 0 & 0 & 0 & 1 \\ 0 & 0 & 1 & 0 \\ 0 & -1 & 0 & 0 \\ 1 & 0 & 0 & 0 \end{bmatrix}, \quad a^2 = \begin{bmatrix} 0 & 0 & -1 & 0 \\ 0 & 0 & 0 & 1 \\ 1 & 0 & 0 & 0 \\ 0 & 1 & 0 & 0 \end{bmatrix}, \quad a^3 = \begin{bmatrix} 0 & 1 & 0 & 0 \\ -1 & 0 & 0 & 0 \\ 0 & 0 & 0 & 1 \\ 0 & 0 & 1 & 0 \end{bmatrix} \quad (4)$$

Matrices $\eta^\mu$ satisfy the relation

$$\eta^\alpha \eta^\beta + \eta^\beta \eta^\alpha = 2g^{\alpha\beta} \quad (5)$$

$$(g^{00} = -g^{11} = -g^{22} = -g^{33} = 1, g^{\mu\nu} = 0 \text{ for } \mu \neq \nu)$$

and one can easily deduce from it

$$a^2 = \frac{e^2}{K^2} A_\alpha \eta^\alpha A_\beta \eta^\beta = \frac{e^2}{K^2} A_\mu A^\mu \quad (6)$$

The Lagrangian density that describes the system *electromagnetic field + real Dirac field* reads

$$L = L_{em} + L_{DI} \quad (7)$$

where

$$L_{em} = \frac{1}{8\pi} \left( -\frac{1}{2} F_{\mu\nu} F^{\mu\nu} + F^2 - G^2 \right) \quad (8)$$

$$G = \partial_\alpha A^\alpha, \quad F = \partial_\alpha C^\alpha \quad (9)$$

and $L_{DI}$ is given by (1).

Since F and G are scalars one can take F=0, G=0.

In this case the Lagrange's equation for $A^\alpha$ is given by

$$\partial_\mu \partial^\mu A^\alpha = 4\pi e \left[ \kappa^2 \overline{\Phi} \eta^\alpha \Phi + \frac{\overline{\Psi} \eta^\alpha \Psi}{1-a^2} - 2\frac{e}{k} \frac{A^\alpha \overline{\Psi}(1-a)\Psi}{(1-a^2)^2} \right] \quad (10)$$

The Lagrange's equation for $\Phi$ reads

$$\partial_\mu \eta^\mu \left[ \frac{(1-a)}{1-a^2} \partial_\nu \eta^\nu \Phi \right] + \kappa^2(1+a)\Phi = 0 \quad (11)$$

The conjugate momenta to $\Phi$ and $\Phi^+$ are respectively

$$\Pi_\Phi = \frac{K}{c} \frac{\overline{\Psi}(1-a)\eta^0}{1-a^2} \quad (12)$$

$$\Pi_{\Phi^+} = \frac{K}{c} \frac{(1-a)\Psi}{1-a^2} \tag{13}$$

Due to the relation

$$\eta^0 \eta^{j+} = \eta^j \eta^0 \tag{14}$$

one obtains

$$\Pi_{\Phi^+} = \Pi_\Phi^{+} \tag{15}$$

The Hamiltonian density that is deduced from (1) reads

$$H = \frac{c^2}{K}\Pi_\Phi \eta^0 (1+a) \Pi_{\Phi^+} - c\Pi_\Phi \partial_j \eta^0 \eta^j \Phi - c\partial_j \overline{\Phi} \eta^j \Pi_{\Phi^+} +$$

$$+ K\kappa^2 \overline{\Phi}(1+a)\Phi \tag{16}$$

Canonical equations of the real Diac field with the electromagnetic interaction are

$$D\Phi - (1+a)\frac{c}{K}\Pi_{\Phi^+} = 0 \tag{17}$$

$$\overline{D\Phi} - \frac{c}{K}\Pi_\Phi \eta^0 (1+a) = 0 \tag{18}$$

$$D\frac{c}{K}\Pi_{\Phi^+} + \kappa^2 (1+a)\Phi = 0 \tag{19}$$

$$\overline{D\frac{c}{K}\Pi_{\Phi^+}} + \kappa^2 \overline{\Phi}(1+a) = 0 \tag{20}$$

Since the equation (18) can be formally obtained from (17) as well as the equation (20) from (19), then it suffices to consider the system of two canonical equations (17) and (19).

## 3. CLASSIFICATION OF THE SOLUTIONS OF CANONICAL EQUATIONS

In order to classify the solutions of the system (17) and (19) one can make use of the system of square matrices 8x8 with real elements that commute with the matrices $\eta^\alpha$ (defined by (4)) and have their inverse matrices. Any matrix Z of this system satisfies the relations:

$$Z\eta^\alpha = \eta^\alpha Z \tag{21}$$

$$Z^2 = -1 \tag{22}$$

$$Z^{-1} = Z^+ \tag{23}$$

After multiplying by $Z\kappa$ the equation (17) from the left one obtains

$$DZ\kappa\Phi - \kappa Z(1+a)\frac{c}{K}\Pi_{\Phi^+} = 0 \tag{17'}$$

where use has been made of (21).

Application of the properties (21) and (22) to the equation (19) gives

$$D\frac{c}{K}\Pi_{\Phi^+} - \kappa Z(1+a)Z\kappa\Phi = 0 \tag{19'}$$

If one subtracts the equation (17') from the equation (19') one obtains

$$D\left(\frac{c}{K}\Pi_{\Phi^+} - Z\kappa\Phi\right) + \kappa Z(1+a)\left(\frac{c}{K}\Pi_{\Phi^+} - Z\kappa\Phi\right) = 0 \tag{24}$$

The addition of the equations (17') and (19') yields

$$D\left(\frac{c}{K}\Pi_{\Phi^+} + Z\kappa\Phi\right) - \kappa Z(1+a)\left(\frac{c}{K}\Pi_{\Phi^+} + Z\kappa\Phi\right) = 0 \tag{25}$$

Let

$$X = \frac{c}{K}\Pi_{\Phi^+} - Z\kappa\Phi \tag{26}$$

$$Y = \frac{c}{K}\Pi_{\Phi^+} + Z\kappa\Phi \tag{27}$$

then the system of equations (24) and (25) becomes

$$DX + \kappa Z(1+a)X = 0 \tag{24'}$$

$$DY - \kappa Z(1+a)Y = 0 \tag{25'}$$

If general solution of partial differential equation (24') is $X_g$ and if general solution of partial differential equation (25') equals $Y_g$ then due to (26) and (27) general solution of the system (17) and (19) reads

$$\left(\frac{c}{K}\Pi_{\Phi^+}\right)_g = \frac{X_g + Y_g}{2} \tag{28}$$

$$(\kappa\Phi)_g = \frac{Z(X_g - Y_g)}{2} \tag{29}$$

Since the equations (24') and (25') are homogeneous partial differential equations then there exists for each of them trivial solution:

$$X = 0 \tag{30}$$
$$Y = 0 \tag{31}$$

If both X and Y vanishes then $\frac{c}{K}\Pi_{\Phi^+} = \kappa\Phi = 0$. This case is of no physical interest.

Physically meaningfull special cases to be considered are:

I) $X = 0, Y = Y_g$ $\quad \frac{c}{K}\Pi_{\Phi^+} = \frac{1}{2}Y_g$ , $\kappa\Phi = -\frac{1}{2}ZY_g$

$$\frac{c}{K}\Pi_{\Phi^+} = Z\kappa\Phi \tag{32}$$

II) $X = X_g, Y = 0$ $\quad \frac{c}{K}\Pi_{\Phi^+} = \frac{1}{2}X_g$ , $\kappa\Phi = \frac{1}{2}ZX_g$

$$\frac{c}{K}\Pi_{\Phi^+} = -Z\kappa\Phi \tag{33}$$

In the case I) the equations (17) and (19) become
$$D\Phi - \kappa(1+a)Z\Phi = 0 \tag{17a}$$
$$DZ\Phi + \kappa(1+a)\Phi = 0 \tag{19a}$$

Note that in this case Lagrange's equation (11) acquires the form of the equation (19a).

If the solutions of (17a) are inserted into (1) one gets
$$(\mathsf{L}_{DI})_I = 0$$

where use has been made of the property $Z^+ = -Z$ that can be derived from (22) and (23).

The current density four-vector is generally given by

$$j^\alpha = -c\frac{\partial \mathsf{L}_{DI}}{\partial A_\alpha} = ce\left[\kappa^2\overline{\Phi}\eta^\alpha\Phi + \frac{\overline{\Psi}\eta^\alpha\Psi}{1-a^2} - 2\frac{e}{K}A^\alpha\frac{\overline{\Psi}(1-a)\Psi}{(1-a^2)^2}\right] \tag{34}$$

In the case I) one obtains

$$\frac{\overline{\Psi}\eta^\alpha\Psi}{1-a^2} = \kappa^2\overline{\Phi}\eta^\alpha\Phi + 2\frac{e}{K}A^\alpha\frac{\kappa^2\overline{\Phi}(1+a)\Phi}{1-a^2}$$

$$\frac{\overline{\Psi}(1-a)\Psi}{(1-a^2)^2} = \frac{\kappa^2 \overline{\Phi}(1+a)\Phi}{1-a^2}$$

and the current-density four-vector becomes

$$j^\alpha = 2ce\kappa^2 \overline{\Phi}\eta^\alpha \Phi \tag{35}$$

Therefore in the case I) the Lagrange's equation for $A^\alpha$ (10) acquires the following form

$$\partial_\mu \partial^\mu A^\alpha = 8\pi e\kappa^2 \overline{\Phi}\eta^\alpha \Phi \tag{36}$$

So in this case the behaviour of the system *electromagnetic field+real Dirac field* is determined by the equations (17a) and (36) (or by the equations (19a) and (36) ). According to (34) the current density four-vector is generally constructed by the real spinors $\Phi, D\Phi$ and the electromagnetic field functions $A^\alpha$. Thus the Lagrangian density (7) and from it derived equations (10),(17),(19) establish in general the theory of nonlinear electrodynamics with real Dirac field. In contrast the equations (17a) and (36) are basic equations of linear electrodynamics with real Dirac field.

In the case II) the equations (17) and (19) become

$$D\Phi + \kappa(1+a)Z\Phi = 0 \tag{17b}$$

$$DZ\Phi - \kappa(1+a)\Phi = 0 \tag{19b}$$

and Lagrange's equation (11) acquires the form of equation (19b). If the solutions of (17b) are inserted into (1) one obtains

$$(\mathcal{L}_{DI})_{II} = 0$$

In the case II) the current density four-vector is also given by (35) and the behaviour of the system *electromagnetic field+real Dirac field* is determined by the equations (17b) and (36) (or by the equations (19b) and (36) ). Therefore the equations (17b) and (36) are also basic equations of linear electrodynamics with real Dirac field. The case II) is equivalent to the case I).

As it is the case in classical theory the equations (17a) and (36) (or the equations (17b) and (36) ) imply the appearance of the self-interaction because particular solution $A_p^\alpha$ of equation (36) is determined by real Dirac field $\Phi$. Effects of the self-interaction are to be reduced to the radiation effects [2].

## 4. CONCLUSIONS

Lagrangian density (1) defines the real Dirac field with the electromagnetic interaction and implies the existence of:

A) nonlinear electrodynamics with real Dirac field
B) linear electrodynamics with real Dirac field

The relations (32) and/or (33) are sufficient conditions for linear electrodynamics with real Dirac field.

References

[1] K.Ljolje , The Dirac Field in Real Domain
[2] S.Botrić and K.Ljolje , On the Self-Interaction in Theory of Real
             Dirac Field